\documentclass[12pt]{article}
\usepackage{latexsym}
\usepackage{amsthm,authblk}
\usepackage{amsmath,amssymb,amsfonts,bm,amsbsy,bbm}
\usepackage{epsfig}
\usepackage{graphicx,rotating,lscape}
\usepackage{verbatim}
\usepackage[authoryear]{natbib}
\usepackage{multirow,color}
\usepackage{geometry}
\usepackage{setspace}
\usepackage{subfigure}
\usepackage{float}
\usepackage{physics}
\usepackage{soul}
\usepackage{url}

\newtheorem{theorem}{Theorem}

\newtheorem{proposition}{Proposition}

\newtheorem{remark}{Remark}

\def\e{{\bf e}}

\def\Q{{\bf Q}}

\def\U{{\bf U}}

\def\V{{\bf V}}

\def\X{{\bf X}}
\def\x{{\bf x}}

\def\uu{{\mathrm{u}}}

\def\P{{\bf P}}

\def\bb{{\boldsymbol\beta}}

\def\bphi{\boldsymbol\phi}

\def\bOmega{{\boldsymbol\Omega}}

\def\0{{\bf 0}}
\def\1{{\bf 1}}
\def\trans{^{\rm T}}

\def\wh{\widehat}
\def\wt{\widetilde}

\def\bse{\begin{eqnarray*}}
	\def\ese{\end{eqnarray*}}
\def\be{\begin{eqnarray}}
	\def\ee{\end{eqnarray}}
\def\bsq{\begin{equation*}}
	\def\esq{\end{equation*}}
\def\bq{\begin{equation}}
	\def\eq{\end{equation}}

\def\sumi{\sum_{i=1}^n}

\def\sumIP1{\sum_{i=1, i\in P_1}^N}

\def\boxit#1{\vbox{\hrule\hbox{\vrule\kern6pt\vbox{\kern6pt#1\kern6pt}\kern6pt\vrule}\hrule}}

\usepackage{tikz}
\usetikzlibrary{positioning,shapes.geometric}

\makeatletter
\newcommand*{\ind}{%
	\mathbin{%
		\mathpalette{\@ind}{}%
	}%
}
\newcommand*{\nind}{%
	\mathbin{
		\mathpalette{\@ind}{\not}
	}%
}
\newcommand*{\@ind}[2]{%
	\sbox0{$#1\perp\m@th$}
	\sbox2{$#1=$}
	\sbox4{$#1\vcenter{}$}
	\rlap{\copy0}
	\dimen@=\dimexpr\ht2-\ht4-.2pt\relax
	\kern\dimen@
	{#2}%
	\kern\dimen@
	\copy0 
}
\makeatother

\title{Efficient and Model-Agnostic Parameter Estimation Under Privacy-Preserving Post-randomization Data}
\author[1]{Qinglong Tian}
\author[2]{Jiwei Zhao}
\affil[1]{University of Waterloo}
\affil[2]{University of Wisconsin-Madison}

\begin{document}

\maketitle

\begin{abstract}
    Protecting individual privacy is crucial when releasing sensitive data for public use. While data de-identification helps, it is not enough. This paper addresses parameter estimation in scenarios where data are perturbed using the Post-Randomization Method (PRAM) to enhance privacy. Existing methods for parameter estimation under PRAM data suffer from limitations like being parameter-specific, model-dependent, and lacking efficiency guarantees. We propose a novel, efficient method that overcomes these limitations. Our method is applicable to general parameters defined through estimating equations and makes no assumptions about the underlying data model.
We further prove that the proposed estimator achieves the semiparametric efficiency bound, making it optimal in terms of asymptotic variance.
\end{abstract}

\newpage

\section{Introduction}

\subsection{Background}

To make data accessible to the public, agencies like the Census Bureau, medical institutes, and law enforcement often release statistical databases.
However, when statistical databases are released publicly, attackers can use them to identify individuals or uncover sensitive information by combining them with other available databases.
For example, \citet{sweeney2001computational} was able to reidentify some cancer patients in an anonymous medical database by matching some information (e.g., gender or zip code) in some named external databases (e.g., voter registration database).
Thus, data de-identification alone is not enough to protect individuals' private information.
In addition to data de-identification, where information like name is removed, statistical disclosure control (SDC) methods aim to prevent the sensitive information from being re-identified in inference attacks.
One can find a general introduction to SDC methodology in \cite{hundepool2012statistical} and \cite{willenborg2012elements}.

Data perturbation is a type of commonly used SDC methods.
It involves intentionally modifying data before publication to safeguard individual privacy.
The core concept is straightforward: sensitive information (e.g., gender, zip code, and date of birth) needs to be altered in a way that prevents linking it to external datasets; thus making inference attacks difficult to implement.
Common approaches include adding simple additive Gaussian noise or utilizing the Laplacian noise mechanism within the differential privacy framework.
We refer to \citet{mivule2012} and \citet{ okkalioglu2015survey} for more details.

This paper considers the problem where we need to perturb categorical variables.
The post-randomization method (PRAM), which is firstly proposed by \cite{gouweleeuw1998post}, is a natural way to adding noise to categorical variables: Suppose there is a sensitive binary variable $Z\in\left\{0,1\right\}$ that we wish to perturb,
the PRAM method performs the perturbation using a known transition matrix, denoted by $\P\in\mathbb{R}^{2\times2}$.
In this example, suppose the transition matrix is defined as
\[
\P=
\begin{bmatrix}
	\Pr(Z^\ast=0|Z=0) & \Pr(Z^\ast=0|Z=1)\\
	\Pr(Z^\ast=1|Z=0) & \Pr(Z^\ast=1|Z=1)
\end{bmatrix}=
\begin{bmatrix}
	0.8 & 0.1\\
	0.2 & 0.9
\end{bmatrix},
\]
where $Z$ is the original unperturbed sensitive variable, and $Z^\ast$ is the perturbed one.
We can see that the PRAM method performs data perturbation probabilistically:
There is a probability ($\Pr(Z^\ast=0|Z=0)=0.8$ or $\Pr(Z^\ast=1|Z=1)=0.9$) that the variable can keep its original value after perturbations; otherwise, its value will be flipped to the opposite category.
Similarly, we can extend the PRAM method to variables with $k\geq3$ categories with a $k\times k$ transition matrix.

The transition matrix $\P$ is crucial in balancing data privacy with data utility.
While all columns in $\P$ must sum to 1, with each entry between 0 and 1, the value along the diagonal (i.e., $\Pr(Z^\ast=i|Z=i)$ for $i=1,\dots,k$) is typically set to be greater than 0.5.
Such a requirement prioritizes preserving the original data up to a certain extent, as overly emphasizing privacy would significantly reduce the usefulness of the published data.
For example, in an extreme case, all entries in $\P$ are set to 0.5 in the binary example, the resulting perturbed variable would contain no usable information.

\subsection{Existing Work and Motivations}

PRAM's probabilistic nature inherently protects against inference attacks.
However, for users who do not have access to the original data, how to use the perturbed data for faithful statistical analysis  poses significant challenges.
Consider a simple example: Suppose we have a logistic regression task with the ``PRAM-ed'' response variable $Y^\ast$ (from the original variable $Y$) and the original covariate $X$, and we are interested in estimating a parameter $\beta$ defined as the solution to an estimating equation $E\left\{U(X,Y;\beta)\right\}=0$ for some known function $U(\cdot)$ (e.g., $\beta$ can be the coefficient of $X$ in logistic regression by choosing $U(\cdot)$ accordingly, see Section~\ref{sec:setting} for more details).
The estimation of $\beta$ is generally biased if we treat the perturbed $Y^\ast$ as if it is the original response $Y$ without making any adjustment.
This is because $\beta^\ast$ (which is the solution to the equation $E\left\{U(X,Y^\ast;\beta^\ast)\right\}=0$) is generally different from $\beta$.
So, even if the estimation method is unbiased given the original data, it becomes biased with PRAM data because we estimate a totally different parameter $\beta^\ast\neq\beta$.

There is some existing work on parameter estimation with PRAM-ed data.
In their seminal works, \cite{gouweleeuw1998post} proposed an unbiased moment estimator for frequency counts, whereas \cite{van2002randomized} studied a general framework to estimate odds ratios.
Provided that a suitable parametric model is available, the EM algorithm \citep{dempster1977maximum} appeared to be a popular choice to adjust for PRAM data.
\cite{van2006estimating} estimated the parameters in a linear regression model when covariates are subject to randomized response.
\cite{woo2012logistic} developed and implemented EM-type algorithms to obtain maximum likelihood estimates in logistic regression models, and \cite{woo2015generalised} further extended the framework to generalized linear models.
In both \cite{woo2012logistic} and \cite{woo2015generalised}, the variables subject to PRAM could be either response, covariate, or both.
However, existing methods have the following major limitations:
\begin{enumerate}
	\item \textbf{Parameter-specificity}: These methods are often designed for specific parameters and may not apply to more general ones.
	\item \textbf{Parametric model dependence}: Many methods rely on specific assumptions about the underlying data model, such as assuming a logistic regression relationship between $p(y|x)$.
	This dependence can make them vulnerable to model misspecification.
	\item \textbf{Limited Optimality}: Existing methods may not offer a guaranteed optimal solution for all situations.
	Specific data characteristics and analysis goals can heavily influence their performance.
\end{enumerate}

\subsection{Related Work}

The PRAM problem is closely related to the label noise problem in the machine learning literature (e.g., \citealt{lawrence2001estimating, pmlr-v38-scott15, li2021provably, pmlr-v202-liu23g, guo2024label}), as well as the misclassification problem in the statistical literature (e.g., \citealt{carroll2006measurement, buonaccorsi2010measurement, yi2017statistical, grace2021likelihood}).
The label noise problem is usually considered in a supervised learning setting, and the goal is to train a classifier using labeled data.
However, we can only observe contaminated label $Y^\ast$ instead of clean label $Y$.
Such a problem is common in real-world applications.
In a medical context, obtaining an accurate gold standard for diagnosis can be challenging due to factors such as cost, time limitations, and ethical considerations. This often necessitates the use of less reliable, ``imperfect'' diagnostic procedures, leading to potential misdiagnoses. Consequently, the labels assigned (healthy or diseased) based on these imperfect methods can introduce noise into the data.

In the context of label noise or misclassification, a common assumption known as class-dependent noise (\citealt{lawrence2001estimating}) states that the probability of a noisy label $Y^\ast$ given the true label $Y$ and the features $\X$ is equal to the probability of the noisy label $Y^\ast$ given only the true label $Y$:
$\Pr(Y^\ast|Y,\X)=\Pr(Y^\ast|Y)$.
Under the class-dependent noise assumption, the label noise and PRAM problems share similar settings. However, a crucial distinction lies in the \textbf{intentionality of the noise}. In label noise scenarios, misclassification occurs unintentionally, arising from various factors such as human error, measurement limitations, or imperfect diagnostic procedures.
Conversely, the noise is \textit{deliberately} introduced through the known transition matrix $\P$ (representing $\Pr(Y^\ast|Y)$) in the PRAM setting.
Unlike the label noise setting, where the transition matrix is typically unknown, the PRAM setting assumes the transition matrix $\P$ is readily available for the general public because $\P$ itself is not sensitive.
This access to the transition matrix allows for different approaches and analyses compared to the unintentional noise encountered in typical label noise problems.

\subsection{Overview}

We propose a novel method for parameter estimation with PRAM data, and we allow the PRAM-ed variables to be either the response or covariate.
We address the aforementioned limitations of existing methods accordingly.
\begin{enumerate}
	\item \textbf{General Parameters}: we consider a parameter $\boldsymbol{\bb}$ defined through an estimating equation $E\left\{U(\X, Y;\boldsymbol{\bb})\right\}=\0$, which is general and covers many commonly used parameters.
	\item \textbf{Model-Agnostic Method}: we do not impose any parametric assumptions on the data model, and our proposed method is free of the problem of model misspecifications.
	\item \textbf{Estimation Efficiency}: We answer the optimality question by proving that our proposed method achieves the semiparametric efficiency bound (\citealt{bickel1993efficient, tsiatis2006semiparametric}).
	In other words, the proposed estimator has the smallest possible asymptotic variance among all the regular asymptotic linear estimators.
\end{enumerate}

The rest of the paper is organized as follows.
Section~\ref{sec:model-dependent} first formally introduces the problem setup and then explains why existing methods rely on parametric assumptions.
Section~\ref{sec:proposed-method} utilizes the findings in Section~\ref{sec:model-dependent} and proposes an efficient and model-agnostic estimator.
Section~\ref{sec:numerical} conducts comprehensive numerical studies to evaluate and compare the proposed method and existing methods.
Section~\ref{sec:conclusion} concludes the paper with discussions on potential future research.

\section{Understanding Model Dependence in Existing Methods}
\label{sec:model-dependent}

\subsection{Problem Setting}
\label{sec:setting}
We start this section by formally introducing the notation used throughout the paper.
We consider a random vector $(Y,Y^\ast,\X)\sim p(y,y^\ast,\x)$, where
\begin{itemize}
	\item $Y$: Represents the original sensitive categorical variable $Y\in\left\{0,\dots,k-1\right\}$.
	\item $Y^\ast$: Represents the PRAM-ed variable, which is the sensitive variable after applying the privacy-preserving transformation.
	\item $\X$: Represents a vector of covariates associated with the variable of interest.
\end{itemize}
Furthermore, we assume that the perturbed variable $Y^\ast$ is independent of $\X$ conditional on $Y$ (e.g., $p(y^\ast|y,\x)=p(y^\ast|y)$) due to the PRAM mechanism.
However, the original $y$ is unobservable; users can only access data from $p(y^\ast,\x)$ and the transition matrix $\P$, whose $(i,j)$th entry is the probability of transforming $Y=j-1$ to $Y^\ast=i-1$: $\Pr(Y^\ast=i-1|Y=j-1)$.
Though we present the method by transforming the response variable $Y$, it is crucial to note that the following discussions and the proposed method also apply seamlessly to transforming sensitive covariates.
We transform $Y$ here solely for illustrative purposes.

We are interested in estimating a parameter $\bb\in\mathbb{R}^d$, which is defined as the solution to the estimating equation $E\left\{\U(Y,\X;\bb)\right\}=\0$ for a known function $\U(\cdot)\in\mathbb{R}^d$.
This general form of parameter definition encompasses various quantities of interest.
For example, by letting $U(y,x;\beta)=y-\beta$, the parameter of interest becomes the mean of the response $\beta=E(y)$.
For binary $y\in\left\{0,1\right\}$ if we let
$$
\U(y,x;\bb)=
\begin{bmatrix}
	y-\mathrm{expit}(\beta_0+\beta_1x)\\
	\left\{y-\mathrm{expit}(\beta_0+\beta_1x)\right\}x
\end{bmatrix},
$$
where $\mathrm{expit}(v)=1/\left\{1+\exp(-v)\right\}$.
Then $\bb\trans=(\beta_0,\beta_1)$ corresponds to the intercept and coefficient of $x$ in a logistic regression model.
For continuous $y$ (suppose $x$ is a categorical sensitive variable), if we let
\[
\U(y,x;\bb)=
\begin{bmatrix}
	y-\beta_0-\beta_1x\\
	(y-\beta_0-\beta_1x)x
\end{bmatrix},
\]
then $\bb\trans=(\beta_0,\beta_1)$ corresponds to the intercept and coefficient of $x$ in a simple linear regression model.
Importantly, we do not make any model assumptions on $p(y,x)$ when defining these parameters.
The definition of the parameter is independent of the model, making it well-defined even if the model (e.g., $p(y|x)$) is misspecified or unknown.
Lastly, the goal is to estimate and perform statistical inference on the parameter $\bb$ given PRAM data $\{(\x_i,y^\ast_i),i=1,\dots,n\}$.

\subsection{Why Existing Methods Are Model-Dependent?}
\label{sec:why-existing-bad}
This section examines the key limitation of existing methods (e.g., \citealt{van2006estimating}, \citealt{woo2012logistic}, \citealt{woo2015generalised}): their reliance on parametric assumptions. 
By analyzing this limitation, we aim to pave the way for introducing our proposed model-agnostic method, which offers greater flexibility and robustness.
The rest of this section provides a high-level examination of existing model-dependent methods.

Due to the absence of the original label $Y$, we cannot use the estimating equation $E\left\{\U(Y,\X;\bb)\right\}=\0$ directly to solve for $\bb$.
However, we can rewrite the estimating equation as
\begin{equation}
	\label{eq:cond-exp}
	E\left\{\U(Y,\X;\bb)\right\}=E\left[E\left\{\U(Y,\X;\bb)|Y^\ast,\X\right\}\right]=0,
\end{equation}
where the conditional expectation $\V(y^\ast,\x)\equiv E\left\{\U(Y,\X;\bb)|Y^\ast=y^\ast,\X=\x\right\}$ in (\ref{eq:cond-exp}) is a function observable variables $y^\ast$ and $\x$.
Therefore, the conditional expectation $\V(y^\ast,\x)$ is crucial for estimating $\bb$ because once we can derive (or at least estimate) $\V(y^\ast,\x)$, we can estimate $\bb$ by solving $\bb$ from the empirical version of (\ref{eq:cond-exp}), which is given by
\begin{equation}
	\label{eq:estimaing-eq}
	\frac{1}{n}\sum_{i=1}^n\V(y_i^\ast,\x_i;\bb)=\0.
\end{equation}

By the definition of conditional expectation, we have
\[
\V(y^\ast,\x;\bb)=E\left\{\U(Y,\X;\bb)|Y^\ast=y^\ast,\X=\x\right\}=\int \U(y,\x;\bb)p(y|y^\ast,\x)dy,
\]
where we can see that $\V(y^\ast,\x;\bb)$ is determined by $p(y|y^\ast,\x)$.
Using the Bayes rule, we can be further write $p(y|y^\ast,\x)$ as
\begin{equation}
	\label{eq:model-V}
	p(y|y^\ast,\x)=\frac{p(y,y^\ast,\x)}{p(y^\ast,\x)}=\frac{p(y^\ast|y,\x)p(y,\x)}{p(y^\ast,\x)}=\frac{p(y^\ast|y)p(y|\x)}{p(y^\ast|\x)}.    
\end{equation}
Equation~(\ref{eq:model-V}) shows that $p(y|y^\ast,\x)$, or equivalently, $\V(y^\ast,\x;\bb)$, is determined by $p(y^\ast|y)$, $p(y|\x)$, and $p(y^\ast|\x)$.
The first one, $p(y^\ast|y)$, is known, so we only need to focus on the latter two: $p(y|\x)$ and $p(y^\ast|\x)$.
These two conditional distributions are closely connected, as we can show that
\begin{equation}
	\label{eq:two-conditional}
	p(y^\ast|\x)=\int p(y^\ast|y) p(y|\x)dy.
\end{equation}
Namely, if we specify $p(y|\x)$, then $p(y^\ast|\x)$ is determined by (\ref{eq:two-conditional}).
On the contrary, under mild conditions, one can also solve $p(y|\x)$ from the integral equation (\ref{eq:two-conditional}) when given $p(y^\ast|\x)$.
To summarize, \textbf{once we put a parametric assumption on $p(y|\x)$ or $p(y^\ast|\x)$}, we can estimate $p(y|y^\ast,\x)$ as well as $\V(y^\ast,\x;\bb)$.
Then by replacing $\V(y^\ast,\x;\bb)$ with the estimated $\widehat{\V}(y^\ast,\x;\bb)$ in (\ref{eq:estimaing-eq}), we can estimate the parameter $\bb$ by solving the equation.
This explain why existing methods hinge on imposing model assumption on $p(y|\x)$ or $p(y^\ast|\x)$.

So far, our analysis highlights the fundamental limitation of existing methods: their dependence on parametric assumptions for estimating $\bb$.
The question is: how to bypass the parametric assumption?
We provide our answer in the next section.

\section{Towards Efficient Estimation: A Model-Agnostic Approach}
\label{sec:proposed-method}

\subsection{All Roads Lead to Rome (But Some Are Better)}

We consider a simple example where the sensitive variable is binary $Y\in\left\{0,1\right\}$.
We try to answer the following question: Given the distribution of the perturbed variable $p(y^\ast)$ and the transition matrix $\P$, how to recover the distribution of the latent $p(y)$?

From a probabilistic point of view, the marginal distributions of $Y$ and $Y^\ast$ are connected through a reversion matrix $\Q_1$ as follows.
\begin{equation}
	\label{eq:probablistic}
	\begin{bmatrix}
		\Pr(Y=0)\\
		\Pr(Y=1)
	\end{bmatrix}=
	\Q_1
	\begin{bmatrix}
		\Pr(Y^\ast=0)\\
		\Pr(Y^\ast=1)
	\end{bmatrix},
\end{equation}
where
\[
\Q_1\equiv
\begin{bmatrix}
	\Pr(Y=0|Y^\ast=0) & \Pr(Y=0|Y^\ast=1)\\
	\Pr(Y=1|Y^\ast=0) & \Pr(Y=1|Y^\ast=1)
\end{bmatrix}.
\]
The problem with this approach is that the reversion matrix $\Q_1$ is not readily available.
But we can compute the reversion matrix $\Q_1$ using the Bayes rule:
\begin{equation}\label{eq:reversion-1}
	\Q_1=
	\begin{bmatrix}
		\dfrac{\strut p_{11}\pi_0}{\strut p_{11}\pi_0+p_{12}\pi_1} & \dfrac{\strut p_{21}\pi_0}{\strut p_{21}\pi_0+p_{22}\pi_1}\\
		\dfrac{\strut p_{12}\pi_1}{\strut p_{11}\pi_0+p_{12}\pi_1} & \dfrac{\strut p_{22}\pi_1}{\strut p_{21}\pi_0+p_{22}\pi_1}
	\end{bmatrix},
\end{equation}
where $p_{ij}$ is the $(i,j)$th entry of the transition matrix $\P$ and $\pi_i\equiv\Pr(Y^\ast=i)$ for $i=0,1$.

Another way to reach the same goal is simply inverting the transition matrix.
By the definition of the transition matrix, we have
\[
\begin{bmatrix}
	\Pr(Y^\ast=0)\\
	\Pr(Y^\ast=1)
\end{bmatrix}
=\P
\begin{bmatrix}
	\Pr(Y=0)\\
	\Pr(Y=1)
\end{bmatrix}.
\]
Therefore, we can recover the marginal distribution of $Y$ in the following way, as long as $\P$ is nonsingular.
\begin{equation}
	\label{eq:inverse-method}
	\begin{bmatrix}
		\Pr(Y=0)\\
		\Pr(Y=1)
	\end{bmatrix}=
	\Q_2
	\begin{bmatrix}
		\Pr(Y^\ast=0)\\
		\Pr(Y^\ast=1)
	\end{bmatrix},
\end{equation}
where the reversion matrix is given by $\Q_2=\P^{-1}$.

While both reversion matrices $\Q_1$ and $\Q_2$ can achieve the same goal, comparing $\Q_1$ and $\Q_2$ reveals their key difference: the first reversion matrix $\Q_1$ depends on the data model (i.e., $\pi_0$ and $\pi_1$) while the second reversion matrix $\Q_2$ does not.
After all, $\Q_1$ contains $\pi_0=\Pr(Y^\ast=0)$ and $\pi_1=\Pr(Y^\ast=1)$ while $\Q_2$ is purely an inverse of a known matrix and non-probabilistic.
Calling $\Pr(Y^\ast=0)$ a model may sound strange, but that is because we consider a simple example.

To further illustrate the distinction between these approaches and lay the groundwork for our proposed method, we consider a general scenario by introducing a covariate vector $\X$.
The task now becomes recovering the joint distribution $p(y,\x)$ using the noisy $p(y^\ast,\x)$ and the transition matrix \P.
Similarly to the previous case in (\ref{eq:probablistic}), we can derive p(y,x) as follows:
\[
\begin{bmatrix}
	p(y=0,\x)\\
	p(y=1,\x)
\end{bmatrix}
=
\Q_1(y^\ast,\x)
\begin{bmatrix}
	p(y^\ast=0,\x)\\
	p(y^\ast=1,\x)
\end{bmatrix},
\]
where the reversion matrix
\[
\Q_1(y^\ast,\x)\equiv\begin{bmatrix}
	p(y=0|y^\ast=0,\x) & p(y=0|y^\ast=1,\x)\\
	p(y=1|y^\ast=0,\x) & p(y=1|y^\ast=1,\x)
\end{bmatrix}
\]
depends on the data model $p(y|y^\ast,\x)$, which is similar to the issue encountered in Section~\ref{sec:why-existing-bad}.
This dependence on the data model underscores the reason why existing methods are model-dependent: they need to estimate $p(y|y^\ast,\x)$.

In contrast, the second approach only requires the known matrix $\Q_2=\P^{-1}$, which is the inverse of the transition matrix.
It can be verified that
\begin{equation}\label{eq:inverse_y_x}
	\begin{bmatrix}
		p(y=0,\x)\\
		p(y=1,\x)
	\end{bmatrix}
	=
	\Q_2
	\begin{bmatrix}
		p(y^\ast=0,\x)\\
		p(y^\ast=1,\x)
	\end{bmatrix}.
\end{equation}
Our discussion above highlights the key advantage of using the reversion matrix $\Q_2=\P^{-1}$: It enables us to bypass the need for parametric assumptions.
Therefore, to answer the question posed at the end of Section~\ref{sec:why-existing-bad} (``how to bypass the parametric assumption?''), our solution is simple: use the inverse of the transition matrix $\P$.

\subsection{A Model-Agnostic Estimator}
\label{sec:model-agnostic-method}
To enhance understanding, in this section, we present our method using a simple example: the parameter of interest $\beta$ is a scalar and the sensitive variable $Y\in\{0,1\}$ is binary.
We will explain the core concept using intuitive language in this section before delving into the formal details in Section~\ref{eq:optimal}.

The parameter of interest $\beta$ is defined as the solution to the estimation equation $E\left\{\uu(Y,\X;\beta)\right\}=0$.
Suppose we have a sample $\left\{(\x_i,y^\ast_i)\right\}$ for $i=1,\dots,n$, then we can rewrite the expectation as
\begin{equation}
	\label{eq:informal}
	\begin{split}
		E&\left\{\uu(Y,\X;\beta)\right\}=\int \uu(\x,y;\beta)p(\x,y)dyd\x=\int\sum_{j=0}^1\uu(\x,y=j;\beta)p(\x,y=j)dyd\x\\
		=&\sum_{i=1}^{n}\sum_{j=0}^1\uu(\x_i,y=j;\beta)p(\x_i,y=j)=\sum_{i=1}^{n}
		\begin{bmatrix}
			\uu(\x_i,y=0;\beta)\\
			\uu(\x_i,y=1;\beta)
		\end{bmatrix}\trans
		\begin{bmatrix}
			p(\x_i,y=0)\\
			p(\x_i,y=1)
		\end{bmatrix}
	\end{split}
\end{equation}
One can understand the third equation in (\ref{eq:informal}) by treating $\x$ as if it is a discrete variable with support $\x\in\{\x_1,\dots,\x_n\}$.
By the relationship in (\ref{eq:inverse_y_x}), we can further write (\ref{eq:informal}) as
\begin{equation}\label{eq:intermidate}
	E\left\{\uu(Y,\X;\beta)\right\}=\sum_{i=1}^n
	\begin{bmatrix}
		\uu(\x_i,y=0;\beta)\\
		\uu(\x_i,y=1;\beta)
	\end{bmatrix}\trans
	\P^{-1}
	\begin{bmatrix}
		p(\x=\x_i,y^\ast=0)\\
		p(\x=\x_i,y^\ast=1)
	\end{bmatrix}.
\end{equation}

Equation~(\ref{eq:intermidate}) implies that $E\left\{\uu(Y,\X;\beta)\right\}$, which is originally an expectation with respect to $(\X,Y)$, can be transformed into an expectation with respect to $(\X,Y^\ast)$ with the help of $\P^{-1}$.
Therefore, we can estimate the expectation in (\ref{eq:intermidate}) using the Monte Carlo method with the sample $\left\{(\x_i,y^\ast_i)\right\}$ for $i=1,\dots,n$, and estimate the parameter $\beta$ by solving the following equation:
\begin{equation}
	\label{eq:empirical-est}
	\widehat{E}\left\{\uu(Y,\X;\beta)\right\}=\frac{1}{n}\sum_{i=1}^n\begin{bmatrix}
		\uu(\x_i,y=0;\beta)\\
		\uu(\x_i,y=1;\beta)
	\end{bmatrix}\trans
	\P^{-1}
	\begin{bmatrix}
		\mathbbm{1}_{y^\ast_i=0}\\
		1-\mathbbm{1}_{y^\ast_i=0}
	\end{bmatrix}=0,
\end{equation}
where the indicator function satisfies $\mathbbm{1}_{y^\ast_i=0}=1$ if $y_i^\ast=0$; otherwise, $\mathbbm{1}_{y^\ast_i=0}=0$.
The empirical estimating equation (\ref{eq:empirical-est}) does not contain any models, thus making our proposed estimator model-agnostic.

\subsection{Theoretical Results}
\label{eq:optimal}

\paragraph{Influence Function}
Firstly, we extend our discussion in Section~\ref{sec:model-agnostic-method} to a general setting where we have a vector of parameters $\bb\in\mathbb{R}^d$ and a multi-class variable $Y\in\left\{0,\dots,K-1\right\}$.
The parameter of interest is defined by the solution of $E\left\{\U(\X,Y;\bb)\right\}=\0\in\mathbb{R}^d$ for some known function $\U(\cdot)\in\mathbb{R}^d$.
Using similar arguments as in Section~\ref{sec:model-agnostic-method}, for $i=1,\dots,K$, the influence function of our proposed estimator $\wh\bb$ is given by
\be
\bphi(y^*=i-1,\x;\bb) &=& \sum_{k=1}^K \U(y=k-1,\x;\bb)\underbrace{q_{k,i}}_{\mbox{element of }\P^{-1}}\label{eq:eifbbpram}\\
&=& \mathbb{U}(\x;\bb) \P^{-1}\e_i,\nonumber
\ee
where $q_{ki}$ is the $(k,i)$th entry of $\P^{-1}$, $\mathbb{U}(\x;\bb)$ is a $d\times K$ matrix with its $k$th column being $\U(y=k-1,\x;\bb)$ such that
\[
\mathbb{U}(\x;\bb)=\begin{bmatrix}
	\U\trans(y=0,\x;\bb)\\
	\vdots\\
	\U\trans(y=K-1,\x;\bb)
\end{bmatrix}\trans,
\]
and $\e_i\in\mathbb{R}^{K}$ is a basis vector whose $i$th entry is 1 while other entries are all 0 for $i=1,\dots,K$.
Based on the influence function in (\ref{eq:eifbbpram}), our proposed estimator $\wh\bb$ given a sample $\left\{(\x_i,y^\ast_i)\right\}$, $i=1,\dots,n$, is the solution to
\be\label{eq:solvebbp}
\frac1n\sumi \sum_{k=1}^K \U(y=k-1,\x_i;\bb)\underbrace{q_{k,y_i^*+1}}_{\mbox{element of }\P^{-1}} = \0.
\ee

\paragraph{Semiparametric Efficiency}

Using the theories of the M-estimator (e.g., \citealt{van2000asymptotic}), the proposed estimator $\widehat\bb$ in (\ref{eq:solvebbp}) is readily consistent and has an asymptotic normal distribution.
Moreover, we will prove in Theorem~\ref{th:efficient-influ} that there are no better alternative estimators regarding estimation efficiency because our proposed $\widehat\bb$ has the smallest possible asymptotic variance (i.e., achieves the semiparametric efficiency bound).

Notably, the proposed estimator $\widehat\bb$ achieves the semiparametric efficiency bound without relying on any additional assumptions.
This is because it operates independent of any models, both parametric and nonparametric. This is a significant advantage, as achieving semiparametric efficiency in other methods often requires some challenging conditions.
For example, literature on double machine learning usually requires that the nonparametric nuisance functions (functions whose specific form is unknown but belongs to a certain class) need to be estimated precisely, typically requiring a convergence rate of $o_p(n^{-1/4})$.
Additionally, for most semiparametric estimator, the parametric part of the model needs to be specified correctly.
We establish the efficiency results in the following theorem.
The proofs are given in the supplementary materials.

\begin{theorem}\label{th:efficient-influ}
	The efficient influence function for estimating $\bb$ is given by
	\[
	\bOmega^{-1}\bphi(y^\ast,\x;\bb),
	\]
	where $\bOmega\in\mathbb{R}^{d\times d}$ is a constant matrix and is defined as $E\left\{\partial\U(\X,Y;\bb)/\partial\bb\trans\right\}$ evaluated at the true value $\bb=\bb_0$ and $\bphi(\cdot)$ is given in (\ref{eq:eifbbpram}).
\end{theorem}

\begin{proposition}
	The proposed efficient $\widehat{\bb}$ has the asymptotic representation
	\[
	\sqrt{n}(\widehat\bb-\bb_0)=\frac{1}{\sqrt{n}}\sum_{i=1}^n\bOmega^{-1}\bphi(y^\ast,\x;\bb_0)+o_p(1)\xrightarrow{d}\mathrm{Norm}\left(\0,\bOmega^{-1}E(\bphi\bphi\trans)(\bOmega^{-1})\trans\right).
	\]
	The proposed estimator $\widehat{\bb}$ achieves the semiparametric efficiency bound (i.e., $\bOmega^{-1}E(\bphi\bphi\trans)(\bOmega^{-1})\trans$), and is semiparametrically efficient.
\end{proposition}

\begin{remark}[Efficiency Loss of $\widehat{\bb}$ Compared to Oracle Estimator $\widehat{\bb}_o$]
	\label{remark:effcomptooracle}
	We define the oracle estimator $\widehat\bb_o$ as the solution of
	\[
	\frac{1}{n}\sum^{n}_{i=1}\U(y_i,\x_i;\bb)=\0.
	\]
	We say $\widehat\bb_o$ is an oracle estimator because it needs the unobservable original label $y_i$.
	The asymptotic representation of the oracle estimator is given as
	\[
	\begin{split}
	&\sqrt{n}(\widehat{\bb}_o-\bb_0)=\frac{1}{\sqrt{n}}\sum_{i=1}^n\bOmega^{-1}\U(y_i,\x_i;\bb_0)+o_p(1)\\
	&\xrightarrow{d}\mathrm{Norm}\left(\0,\bOmega^{-1}E(\U\U\trans)(\bOmega^{-1})\trans\right).
	\end{split}
	\]
	Noticing that $E\left\{\bphi(y^\ast,\x;\bb)|Y=y,\X=\x\right\}=\U(\x,y;\bb)$, one can write
	\[
	\begin{split}
	\bOmega^{-1}E(\bphi\bphi\trans)(\bOmega^{-1})\trans
	=&\bOmega^{-1}E(\U\U\trans)(\bOmega^{-1})\trans\\
	&+\bOmega^{-1}E\left\{(\bphi-\U)(\bphi-\U)\trans\right\}(\bOmega^{-1})\trans.
	\end{split}
	\]
	Therefore, $\bOmega^{-1}E\left\{(\bphi-\U)(\bphi-\U)\trans\right\}(\bOmega^{-1})\trans$ is the efficiency loss of $\widehat{\bb}$ compared to the oracle estimator $\widehat{\bb}_o$.
	It is also the price to pay for preserving privacy.
	Considering a special case where $\U(\cdot)$ does not contain $Y$, then from (\ref{eq:eifbbpram}), we have $\bphi=\U$, and there is no efficiency loss because $Y$ is not involved.
\end{remark}

\paragraph{Statistical Inference}
Finally, to preserve the model-agnostic feature of $\wh\bb$ when conducting statistical inference, we propose to use the resampling method by \citet{jin2001simple} to estimate the asymptotic variance of $\wh\bb$.
The details of this method have been carefully studied in \cite{jin2001simple}, so we only briefly describe the procedure here.
Note that $\wh\bb$ is the solution of
$n^{-1}\sumi \bphi(y^\ast_i,\x_i;\bb) = \0$,
where the randomness of $\wh\bb$ comes from different realizations of the data sample $\left\{\left(y_i^\ast,\x_i\right)\right\}$ for $i=1,\dots,n$.
Now, conditional on one observed sample $\left\{\left(y_i^\ast,\x_i\right)\right\}$ for $i=1,\dots,n$, we denote the observed estimate of $\bb$ as $\wh\bb$ (which is fixed now).
Let $L_i,i=1,\ldots,n$, be $n$ independent and identically distributed copies of a nonnegative, completely known random variable $L$ with mean one and variance one (e.g., $L\sim\exp(1)$).
Then, we denote $\wt\bb$ as the solution of
$
n^{-1}\sumi L_i \bphi(y^\ast_i, \x_i;\bb) = \0,
$
where only $L_i$'s are random in the above estimating equation.
It can be shown that $\sqrt{n}(\wt\bb-\wh\bb)$ (when conditional on the sample) shares the same asymptotic distribution as $\sqrt{n}(\wh\bb- \bb_0)$.
In practice, conditional on $(y_i^*, \x_i), i=1,\ldots,n$, and $\wh\bb$, the distribution of $\wt\bb$ can be estimated by generating a large number, say, $M$, of random samples $L_i, i=1,\ldots,n$.
Denote the solution in the $m$th replication as $\wt\bb_{m}$, then the asymptotic variance of $\wh\bb$ can be estimated by the sample covariance matrix constructed from $\wt\bb_{m}, m=1,\ldots, M$.

\section{Numerical Studies}
\label{sec:numerical}

\subsection{Simulation Studies}

We conduct comprehensive simulation studies to investigate the proposed method's finite sample performance and its comparison to some existing methods.
In Simulation A1, we consider a binary $Y\in\left\{0,1\right\}$ and use the logistic regression model $\Pr(y=1|x)=\mathrm{expit}(\wt\beta_0+\wt\beta_1x)$ with $\wt\beta_0=-1$ and $\wt\beta_1=1.5$.
The marginal distribution of $X$ is chosen as $\mathrm{Norm}(0.5, 1)$.
The response variable $Y$ is subject to PRAM.
We choose three different transition matrices in PRAM as $\Pr(Y^\ast=0|Y=0)=\Pr(Y^\ast=1|Y=1)\in\left\{0.75,0.85,0.95\right\}$ and we vary the sample size $n\in\left\{1000,1200,1400,1600,1800,2000\right\}$.
The parameters of interest $\bb=(\beta_1,\beta_2)\trans$ are defined through the solution of the following estimating equation:
\[
E
\begin{bmatrix}
	Y-\mathrm{expit}(\beta_1+\beta_2X)\\
	\left\{Y-\mathrm{expit}(\beta_1+\beta_2X)\right\}X
\end{bmatrix}=\0.
\]
Such a choice of $\bb$ is equivalent to estimating the coefficients from a logistic regression model so that, immediately, we have $\beta_1=\wt\beta_1$ and $\beta_2=\wt\beta_2$; however, the estimating equation itself does not imply any model assumptions.
Using the same way for generating $(y,\x)$ as in Simulation~A1, we conduct Simulation~A2, where we fix the sample size at $n=1000$.
However, we do not require $p_{00}\equiv\Pr(Y^\ast=0|Y=0)=\Pr(Y^\ast=1|Y=1)\equiv p_{11}$ for the transition matrix; we vary $p_{00}$ and $p_{11}$ between $0.75$ and $0.95$ with increment of $0.01$.

In Simulation~B, we consider the situation that the sensitive variable is a covariate, rather than the response variable.
The data-generating mechanism is given by $Y|X=x\sim\mathrm{Norm}(\wt\beta_{1}+\wt\beta_2x,1)$ and $\Pr(X=1)=0.5$ for binary $X\in\left\{0,1\right\}$.
We set $\wt\beta_1=-1$ and $\wt\beta_2=1$.
The parameters of interest $\bb=(\beta_1,\beta_2)$ are the coefficients of a simple linear regression model
\[
E
\begin{bmatrix}
	Y-\beta_1-\beta_2X\\
	(Y-\beta_1-\beta_2X)X
\end{bmatrix}
=\0.
\]
Again, the definition of parameters does not depend on parametric model assumptions, but we align the parameter definition with the data-generating mechanism so that we can obtain the true parameter values $\beta_1=\wt\beta_1$ and $\beta_2=\wt\beta_2$.
Similar to Simulations~A1 and A2, in Simulation~B1, we let the sample size vary $n\in\left\{1000,1200,1400,1600,1800,2000\right\}$ and use $p_{11}\equiv\Pr(X^\ast=1|X=1)=\Pr(X^\ast=0|X=0)\equiv p_{00}\in\left\{0.75,0.85,0.95\right\}$ as the transition probabilities; while in Simulation~B2, we fix $n$ at $n=1000$ and allow $p_{11}$ and $p_{00}$ to vary freely between $0.75$ and $0.95$ with increment of $0.01$.
We run $2000$ Monte Carlo replicates for the following methods in all the aforementioned simulations.
\begin{enumerate}
	\item The \textbf{proposed estimator} $\wh\bb$ in (\ref{eq:solvebbp}).
	\item The \textbf{oracle estimator} $\wh\bb_{o}$: using the un-perturbed original sensitive variable.
	\item The \textbf{naive estimator} $\wh\bb_{b}$: treat the perturbed variable as the original one; this estimator is generally biased.
	\item \textbf{Model-dependent estimator} $\wh\bb_{m1}$ (Model~1): We model (\ref{eq:model-V}) using a parametric model.
	\item \textbf{Model-dependent estimator} $\wh\bb_{m2}$ (Model~2): We model (\ref{eq:model-V}) using a parametric model, but the model used is relatively worse than the one used in $\wh\bb_{m1}$.
\end{enumerate}

The purpose of introducing $\wh\bb_{m1}$ and $\wh\bb_{m2}$ is to investigate the effects of the degree of model misspecification on the estimations.
In Simulations~A (resp., B), we model $p(y^\ast|y,x)$ (resp., $p(x^\ast|x,y)$) using a logistic regression for $\wh\bb_{m1}$ while we force the intercept of the logistic regression to be $0$ in $\wh\bb_{m2}$.
The motivations for such a setting are:
\begin{itemize}
	\item Neither the model in $\wh\bb_{m1}$ or $\wh\bb_{m2}$ is correct; but the parametric model used in $\wh\bb_{m1}$ provides a better fit than that used in $\wh\bb_{m2}$.
	In other words, the degree of model misspecification is worse in $\wh\bb_{m2}$
	\item By comparing $\wh\bb_{m1}$ and $\wh\bb$ as well as $\wh\bb_{m2}$ and $\wh\bb$, we can investigate the benefits brought by our model-agnostic method.
	\item By comparing $\wh\bb_{m1}$ and $\wh\bb_{m2}$,  we can investigate the severity of model misspecification on the estimation.
	It is worth noting that $\wh\bb_{m2}$ does \textit{not} have any practical meanings, nor do we want to establish the superiority of our proposed method by comparing it with $\wh\bb_{m2}$.
	\item We do not include the estimator where $p(y^\ast|y,x)$ (or $p(x^\ast|x,y)$) is correctly specified.
	This is because using the correct model needs oracle information, and we consider the semiparametric setting where the model $p(y^\ast|y,x)$ is unknown (which mimics the real-world application scenario).
\end{itemize}

\begin{center}
	\begin{figure}[th!]
		\centering
		\includegraphics[width=0.9\linewidth]{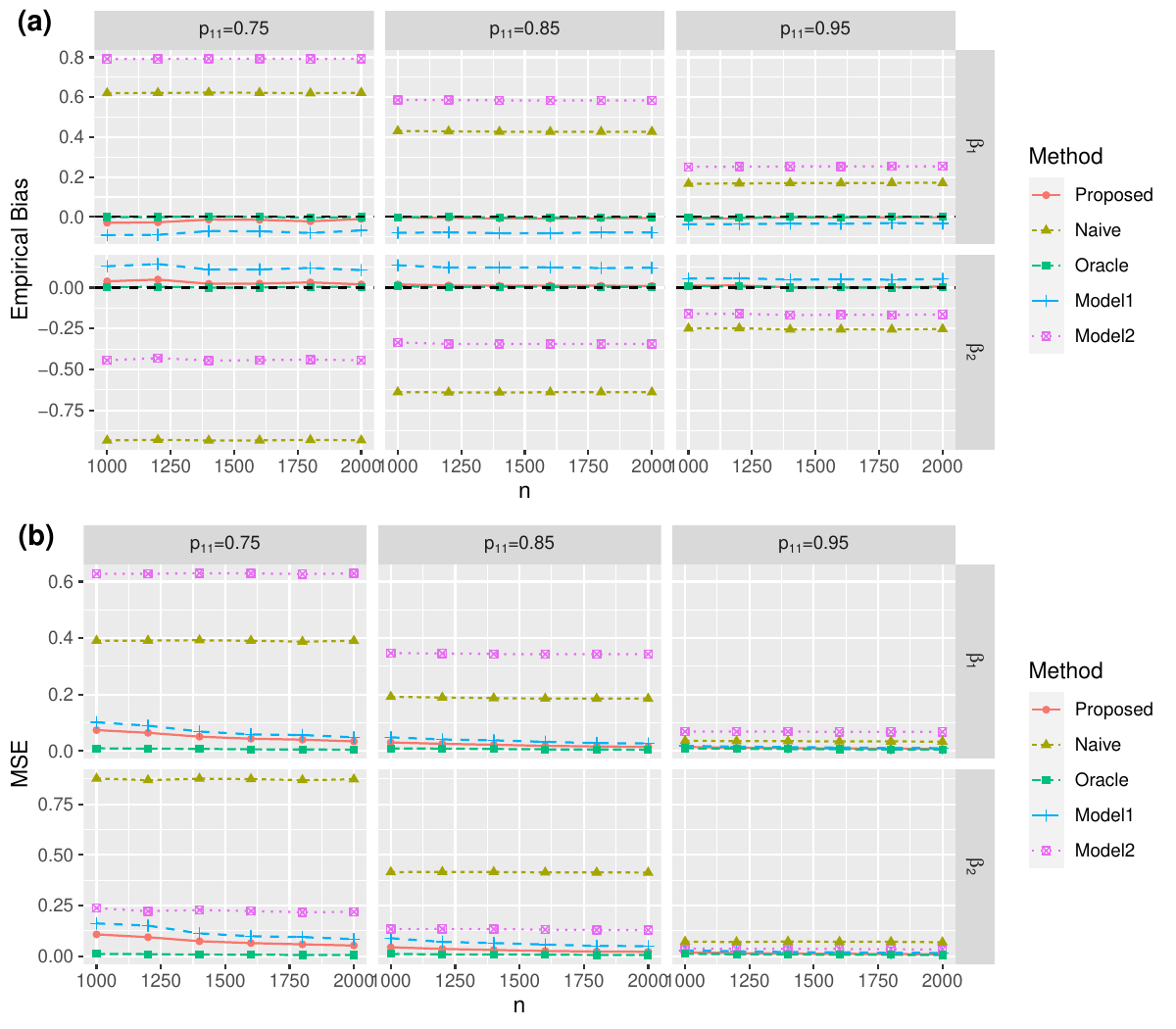}
		\caption{Simulation A1: empirical bias (upper) and MSE (lower) of five estimators.}
		\label{fig:bias1}
	\end{figure}
\end{center}

Firstly, Figure~\ref{fig:bias1}(a) summarizes the empirical bias results in Simulation A1.
In general, the oracle and proposed estimators $\wh\bb_o$ and $\wh\bb$ almost always have no bias, whereas $\wh\bb_{m1}$ has a small bias; however, estimators $\wh\bb_b$ and $\wh\bb_{m2}$ have huge biases.
These noticeable biases reduce when $p_{11}=p_{00}$ increase from $0.75$ to $0.95$ but do not diminish when the sample size $n$ increases.
The empirical bias results in Simulation B1 are very similar, so they are omitted here (contained in the supplementary materials).

We then compare the mean squared error (MSE) across these estimators and report the results in Simulation A1 in Figure~\ref{fig:bias1}(b).
The results in Simulation B1 are similar and omitted here (can be found in the supplementary materials).
Similarly, the naive and Model~2 estimators $\wh\bb_b$ and $\wh\bb_{m2}$ perform badly.
Among the three estimators $\wh\bb_o$, $\wh\bb$, and $\wh\bb_{m1}$, the oracle estimator $\wh\bb_o$ has the smallest MSE while $\wh\bb_{m1}$ has the largest, with our proposed estimator $\wh\bb$ in between, which matches our theoretical results.
The efficiency loss of $\wh\bb$ over $\wh\bb_o$ has been demonstrated in Remark~\ref{remark:effcomptooracle}, which is the price to pay for protecting private information.

\begin{center}
	\begin{figure}[t!]
		\centering
		\includegraphics[width=0.9\linewidth]{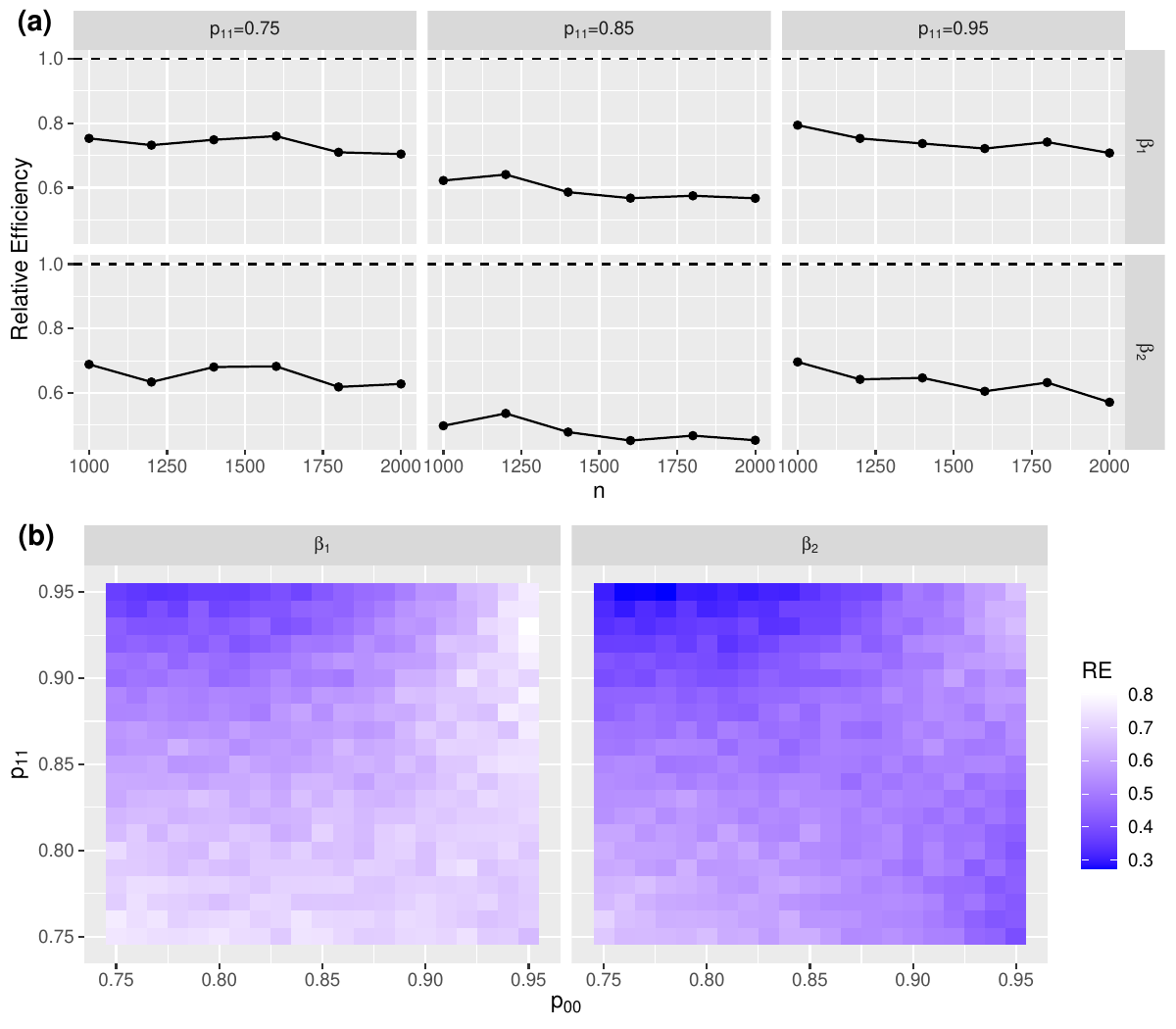}
		\caption{Simulation A1 (upper) and Simulation A2 (lower): Relative efficiency of $\wh\bb$ to $\wh\bb_{m1}$ (relative efficiency smaller than 1 indicates $\wh\bb$ is more efficient).
		}
		\label{fig:re11}
	\end{figure}
\end{center}

\begin{center}
	\begin{figure}[th!]
		\centering
		\includegraphics[width=0.9\linewidth]{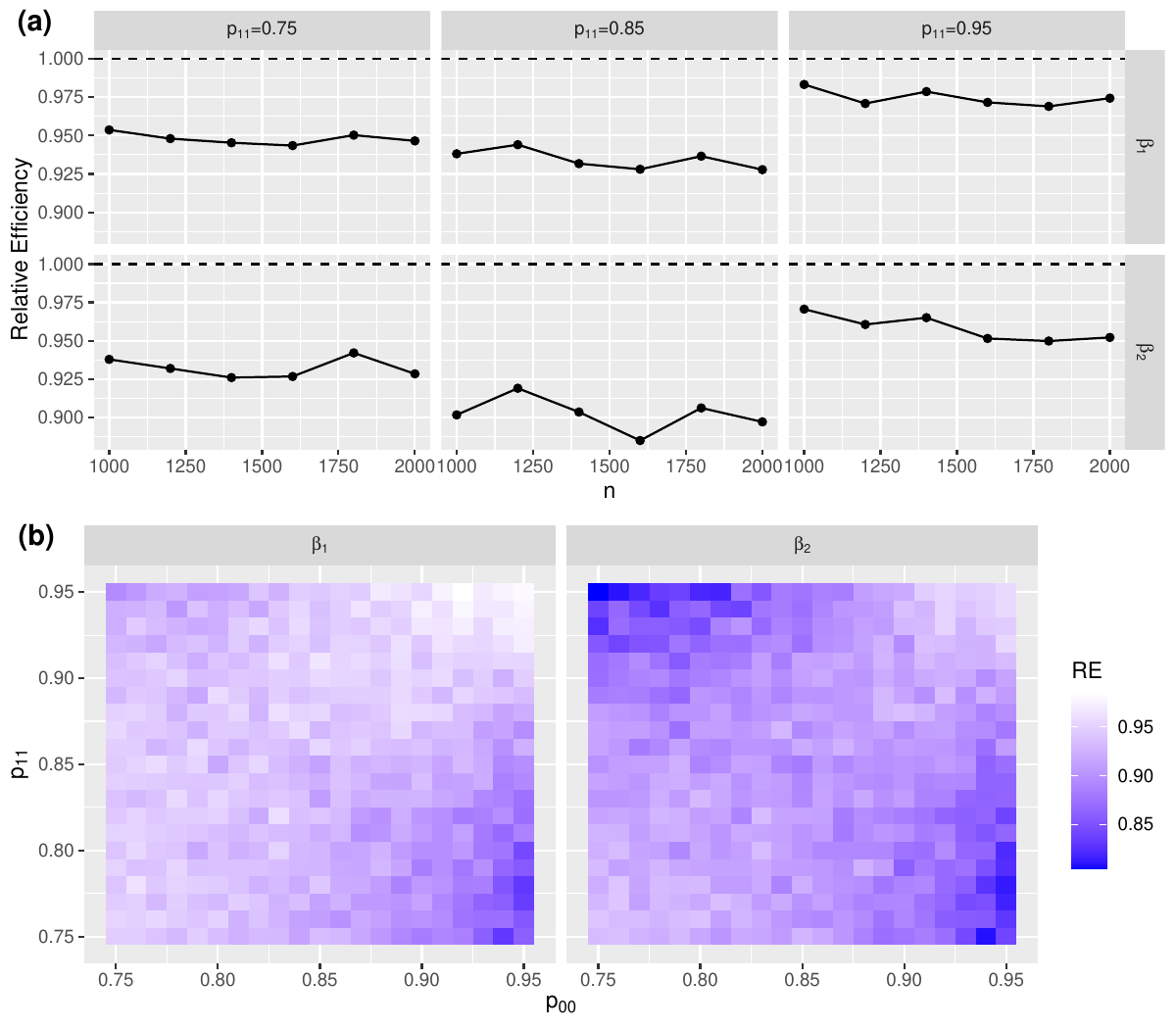}
		\caption{Simulation B1 (upper) and Simulation B2 (lower): Relative efficiency of $\wh\bb_p$ to $\wh\bb_{m1}$ (relative efficiency smaller than 1 indicates $\wh\bb_p$ is more efficient).}
		\label{fig:re21}
	\end{figure}
\end{center}

Next, we compare the estimation efficiency of $\wh\bb$ and $\wh\bb_{m1}$ by assessing the relative efficiency (RE) of $\wh\bb$ to $\widehat{\bb}_{m1}$, which is defined as
$$
\mathrm{RE}(\wh\bb, \widehat{\bb}_{m1})={\mathrm{MSE}(\wh\bb)}/{\mathrm{MSE}(\wh\bb_{m1})}.
$$
If $\mathrm{RE}(\wh\bb, \widehat{\bb}_{m1})<1$, then $\wh\bb$ is preferable than $\wh\bb_{m1}$, and the smaller $\mathrm{RE}(\wh\bb_p, \widehat{\bb}_{m1})$ is, the more efficient $\wh\bb$ is.
Our results in Figure~\ref{fig:re11} show that the proposed estimator $\wh\bb$ is always more efficient than the estimator $\wh\bb_{m1}$ in every situation we consider.
The RE in simulation A1 is about 50\% to 80\% (see Figure~\ref{fig:re11}(a)) while the RE in simulation B1 is about 90\% to 97.5\% (see Figure~\ref{fig:re21}(a)).
When we vary $p_{00}$ and $p_{11}$ in a wider interval $[0.75, 0.95]$, the RE in simulation A2 may range from 30\% to 80\% (see Figure~\ref{fig:re11}(b)) and the RE in simulation B2 can be as small as 80\% (see Figure~\ref{fig:re21}(b)).
Interestingly, in simulations A2 and B2, the area where $\wh\bb_p$ is more efficient is in either the top-left or bottom-right corners.

Lastly, we thoroughly report our proposed estimator $\wh\bb$'s estimation and inference results.
Table~\ref{tab:proposed-logistic} is for simulation A1, and a similar table for simulation B1 is in the supplementary materials.
In Table~\ref{tab:proposed-logistic}, we summarize the empirical bias (Bias) (sample bias across 2000 replicates), the empirical standard deviation (SD) (sample standard deviation across 2000 replicates), the estimated standard error (SE) (average across 2000 estimated standard deviations, with each computed using 500 perturbed samples), and the coverage probability (CP) of the equal-sided $95\%$ confidence interval.
Apparently, the estimated standard error matches the empirical standard deviation closely, and the coverage probability is close to the nominal level $95\%$.

\begin{center}
	\begin{table}[th!]
		\centering
		\begin{tabular}{lccccccccc}
			\hline
			\multirow{2}{*}{$p_{11}\!=\!p_{00}$} &
			\multirow{2}{*}{$n$} &
			\multicolumn{2}{c}{Bias} &
			\multicolumn{2}{c}{SD} &
			\multicolumn{2}{c}{SE} &
			\multicolumn{2}{c}{CP} \\ \cline{3-10}
			&
			&
			$\beta_1$ &
			\multicolumn{1}{c}{$\beta_2$} &
			$\beta_1$ &
			\multicolumn{1}{c}{$\beta_2$} &
			$\beta_1$ &
			\multicolumn{1}{c}{$\beta_2$} &
			$\beta_1$ &
			$\beta_2$ \\ \hline
			\multirow{6}{*}{$0.75$} & 1000 & -0.030 & 0.038 & 0.270 & 0.325 & 0.283 & 0.351 & 0.968 & 0.960 \\
			& 1200 & -0.027 & 0.049 & 0.252 & 0.302 & 0.258 & 0.320 & 0.962 & 0.960 \\
			& 1400 & -0.013 & 0.023 & 0.224 & 0.270 & 0.233 & 0.287 & 0.959 & 0.960 \\
			& 1600 & -0.015 & 0.024 & 0.208 & 0.252 & 0.217 & 0.268 & 0.960 & 0.957 \\
			& 1800 & -0.022 & 0.032 & 0.199 & 0.239 & 0.205 & 0.252 & 0.958 & 0.960 \\
			& 2000 & -0.010 & 0.020 & 0.184 & 0.228 & 0.192 & 0.235 & 0.967 & 0.959 \\ \hline
			\multirow{6}{*}{$0.85$} & 1000 & -0.003 & 0.019 & 0.173 & 0.208 & 0.177 & 0.213 & 0.957 & 0.958 \\
			& 1200 & -0.005 & 0.013 & 0.159 & 0.188 & 0.161 & 0.193 & 0.960 & 0.962 \\
			& 1400 & -0.007 & 0.011 & 0.148 & 0.174 & 0.149 & 0.178 & 0.957 & 0.955 \\
			& 1600 & -0.008 & 0.013 & 0.134 & 0.160 & 0.139 & 0.166 & 0.963 & 0.967 \\
			& 1800 & -0.006 & 0.011 & 0.126 & 0.150 & 0.130 & 0.156 & 0.960 & 0.965 \\
			& 2000 & -0.005 & 0.011 & 0.121 & 0.146 & 0.123 & 0.147 & 0.954 & 0.954 \\ \hline
			\multirow{6}{*}{$0.95$} & 1000 & -0.008 & 0.012 & 0.119 & 0.136 & 0.119 & 0.135 & 0.953 & 0.953 \\
			& 1200 & -0.006 & 0.013 & 0.107 & 0.122 & 0.108 & 0.123 & 0.958 & 0.955 \\
			& 1400 & -0.003 & 0.002 & 0.100 & 0.112 & 0.100 & 0.113 & 0.951 & 0.955 \\
			& 1600 & -0.003 & 0.003 & 0.093 & 0.106 & 0.094 & 0.106 & 0.947 & 0.946 \\
			& 1800 & -0.001 & 0.003 & 0.088 & 0.101 & 0.088 & 0.099 & 0.949 & 0.952 \\
			& 2000 & -0.001 & 0.006 & 0.084 & 0.094 & 0.083 & 0.095 & 0.954 & 0.949 \\ \hline
		\end{tabular}
		\caption{Simulation A1: estimation results and inference results of the proposed estimator $\wh\bb_p$.
		}
		\label{tab:proposed-logistic}
	\end{table}
\end{center}

\subsection{Real Data Application}

To evaluate our method's performance in a practical setting, we applied it to a real dataset from the Korean Labor \& Income Panel Study (dataset provided in the supplementary materials).
This dataset includes information on $n=2505$ regular wage earners for the year 2005. We focused on modeling the average monthly income (denoted by Y as a continuous variable) as a function of three demographic covariates: age ($X_1$, also continuous), education level ($X_2$ as a binary variable, where 1 indicates education beyond high school and 0 otherwise), and gender ($X_3$ as a binary variable, where 1 indicates female and 0 indicates male).
Both income and age were standardized before analysis.

We focus on modeling the average monthly income ($Y$) as a linear function of the three demographic covariates: age ($X_1$), education level ($X_2$), and gender ($X_3$).

The parameters of interest are denoted by the vector $\bb=(\beta_0,\beta_1,\beta_2,\beta_3)\trans$ in our data analysis.
These parameters are defined as the solution to the following estimating equation.
\[
E\left\{(Y-\beta_0-\beta_1X_1-\beta_2X_2-\beta_3X_3)
\begin{bmatrix}
	1 &
	X_1 &
	X_2 &
	X_3
\end{bmatrix}\trans
\right\}=\0
\]
In other words, the parameters are the minimizers of the mean least square error
$$
E\left\{(Y-\beta_0-\beta_1X_1-\beta_2X_2-\beta_3X_3)^2\right\}.
$$
However, the true value of $\bb$ is unknown as we only have a limited sample of data. To assess estimators' performance, we compare them to an oracle estimator (denoted by  $\wh\bb_o$) obtained using the ordinary least squares method, assuming access to all the original variables.

We perform the PRAM procedure on the education level variable $X_2$ to generate a perturbed version $X_2^\ast$.
The transition probabilities controlling the amount of noise introduced during perturbation were set to $p=\Pr(X_2^\ast=0|X_2=0)=\Pr(X_2^\ast=1|X_2=1)\in\left\{0.75, 0.85, 0.95\right\}$.
For each transition probability value, we compare the performance of four estimators
\begin{enumerate}
	\item \textbf{Proposed Estimator} $\wh\bb$.
	\item \textbf{Naive Estimator} $\wh\bb_b$: This estimator treats the perturbed variable $X_2^\ast$ as the original $X_2$, ignoring the impact of PRAM.
	\item \textbf{Model-Dependent Estimator} $\wh\bb_{m1}$ (Model~1): This method uses logistic regression to model $p(x_2^\ast|y,x_1,x_3)$.
	\item \textbf{Model-Dependent Estimator} $\wh\bb_{m2}$ (Model~2): This method uses probit regression for the same purpose as Model 1.
\end{enumerate}
We estimate the standard errors for each method using the resampling approach described in Section~\ref{eq:optimal}.
To evaluate estimator performance, we calculate two metrics:
\begin{itemize}
\item \textbf{Bias}: The difference between the estimated value and the oracle estimator.
\item \textbf{Root Mean Square Error (rMSE)}:This combines the bias and standard error to provide a more comprehensive measure of estimation accuracy: $\text{rMSE}=\sqrt{\text{Bias}^2+\text{SE}^2}$.
\end{itemize}
A lower bias and/or rMSE indicate a better estimator.

The results are summarized in Table~\ref{tab:app1}.
Our proposed estimator consistently achieves the smallest bias and rMSE across all perturbation levels, demonstrating its superiority when compared to the other methods.
This implies that the proposed estimator provides estimates closest to the true values, even when data is perturbed for privacy preservation.
The performance of the other three estimators varies.
While model-dependent methods generally outperform the naive estimator, they can also perform worse in some cases (particularly with low perturbation levels).

\begin{table}[htbp]
	\centering
	\singlespacing
	\footnotesize
	\scalebox{0.82}{
		\begin{tabular}{llcccc}
		\hline
		Parameter & \multicolumn{1}{c}{Method} & \multicolumn{1}{c}{Measure} & \multicolumn{1}{c}{$p=0.75$} & \multicolumn{1}{c}{$p=0.85$} & \multicolumn{1}{c}{$p=0.95$} \\\hline
		\multirow{13}{*}{Intercept} & Oracle & Estimate (SE) &  & 0.410 (0.043) & \\
		\cline{2-6}
		& \multirow{3}{*}{Proposed} & Estimate (SE) & 0.370 (0.062) & 0.411 (0.051) & 0.388 (0.041) \\
		& & ``Bias'' & -0.040 & 0.001 & -0.022 \\
		& & ``rMSE'' & 0.074 & 0.051 & 0.047\\
		\cline{2-6}
		& \multirow{3}{*}{Model1} & Estimate (SE) & 0.317 (0.034) & 0.332 (0.034) & 0.281 (0.031) \\
		& & ``Bias'' & -0.093 & -0.078 & -0.129\\
		& & ``rMSE'' & 0.099 & 0.085 & 0.133\\
		\cline{2-6}
		& \multirow{3}{*}{Model2} & Estimate (SE) & 0.235 (0.038) & 0.267 (0.041) & 0.293 (0.030) \\
		& & ``Bias'' & -0.175 & -0.143 & -0.117\\
		& & ``rMSE'' & 0.179 & 0.149 & 0.121\\
		\cline{2-6}
		& \multirow{3}{*}{Naive} & Estimate (SE) & 0.651 (0.042) & 0.583 (0.043) & 0.454 (0.043) \\
		& & ``Bias'' & 0.242 & 0.173 & 0.044\\
		& & ``rMSE'' & 0.246 & 0.178 & 0.062\\ \hline
		\multirow{13}{*}{Age} & Oracle & Estimate (SE) &  & 0.396 (0.038) &  \\
		\cline{2-6}
		& \multirow{3}{*}{Proposed} & Estimate (SE) & 0.423 (0.049) & 0.404 (0.045) & 0.411 (0.039) \\
		& & ``Bias'' & 0.027 & 0.008 & 0.015\\
		& & ``rMSE'' & 0.056 & 0.046 & 0.042\\
		\cline{2-6}
		& \multirow{3}{*}{Model1} & Estimate (SE) & 0.544 (0.034) & 0.524 (0.036) & 0.538 (0.032) \\
		& & ``Bias'' & 0.148 & 0.128 & 0.142\\
		& & ``rMSE'' & 0.152 & 0.133 & 0.146\\
		\cline{2-6}
		& \multirow{3}{*}{Model2} & Estimate (SE) & 0.609 (0.038) & 0.584 (0.041) & 0.519 (0.032) \\
		& & ``Bias'' & 0.213 & 0.188 & 0.123\\
		& & ``rMSE'' & 0.216 & 0.192 & 0.127\\
		\cline{2-6}
		& \multirow{3}{*}{Naive} & Estimate (SE) & 0.242 (0.038) & 0.287 (0.038) & 0.366 (0.038) \\
		& & ``Bias'' & -0.154 & -0.109 & -0.030\\
		& & ``rMSE'' & 0.159 & 0.115 & 0.048\\ \hline
		\multirow{13}{*}{Edu (PRAM-ed)} & Oracle & Estimate (SE) &  & 0.342 (0.019) & \\
		\cline{2-6}
		& \multirow{3}{*}{Proposed} & Estimate (SE) & 0.371 (0.042) & 0.335 (0.032) & 0.363 (0.021) \\
		& & ``Bias'' & 0.029 & -0.007 & 0.021\\
		& & ``rMSE'' & 0.051 & 0.032 & 0.030\\
		\cline{2-6}
		& \multirow{3}{*}{Model1} & Estimate (SE) & 0.258 (0.010) & 0.274 (0.014) & 0.339 (0.013) \\
		& & ``Bias'' & -0.084 & -0.068 & -0.003\\
		& & ``rMSE'' & 0.085 & 0.069 & 0.013\\
		\cline{2-6}
		& \multirow{3}{*}{Model2} & Estimate (SE) & 0.277 (0.011) & 0.296 (0.013) & 0.349 (0.014) \\
		& & ``Bias'' & -0.065 & -0.046 & 0.007\\
		& & ``rMSE'' & 0.066 & 0.048 & 0.016\\
		\cline{2-6}
		& \multirow{3}{*}{Naive} & Estimate (SE) & 0.170 (0.019) & 0.220 (0.020) & 0.320 (0.019) \\
		& & ``Bias'' & -0.172 & -0.122 & -0.020\\
		& & ``rMSE'' & 0.173 & 0.124 & 0.028\\ \hline
		\multirow{13}{*}{Gender} & Oracle & Estimate (SE) & & -0.290 (0.019) &  \\
		\cline{2-6}
		& \multirow{3}{*}{Proposed} & Estimate (SE) & -0.282 (0.022) & -0.296 (0.018) & -0.288 (0.017) \\
		& & ``Bias'' & 0.008 & -0.006 & 0.002\\
		& & ``rMSE'' & 0.023 & 0.019 & 0.017\\
		\cline{2-6}
		& \multirow{3}{*}{Model1} & Estimate (SE) & -0.275 (0.015) & -0.293 (0.015) & -0.281 (0.014) \\
		& & ``Bias'' & 0.015 & -0.003 & 0.009\\
		& & ``rMSE'' & 0.021 & 0.015 & 0.017\\
		\cline{2-6}
		& \multirow{3}{*}{Model2} & Estimate (SE) & -0.261 (0.014) & -0.289 (0.016) & -0.285 (0.015) \\
		& & ``Bias'' & 0.029 & 0.001 & 0.005\\
		& & ``rMSE'' & 0.032 & 0.016 & 0.016\\
		\cline{2-6}
		& \multirow{3}{*}{Naive} & Estimate (SE) & -0.316 (0.020) & -0.312 (0.020) & -0.296 (0.020) \\
		& & ``Bias'' & -0.026 & -0.022 & -0.006\\
		& & ``rMSE'' & 0.033 & 0.030 & 0.021\\ \hline	\end{tabular}}
	\caption{Real data application: the parameter estimates and standard errors (in parentheses) of all five estimators, as well as ``Bias'' and ``rMSE'' (both compared to the oracle estimator $\wh\bb_o$) of all other four estimators: $\wh\bb$ (Proposed), $\wh\bb_{m1}$ (Model1), $\wh\bb_{m2}$ (Model2), and $\wh\bb_b$ (Naive).}
	\label{tab:app1}
\end{table}

\section{Conluding Remarks}
\label{sec:conclusion}

This paper proposes a novel method for efficient and model-agnostic parameter estimation with data perturbed using the PRAM method for privacy preservation.
Our estimator offers significant advantages over existing methods by overcoming their limitations of parameter-specificity and model dependence. Notably, we prove that the proposed estimator achieves the semiparametric efficiency bound. In simpler terms, this implies that our method offers the best possible accuracy among all estimators when the true data distribution is unknown, which is almost always true in real-world applications.
Looking towards future research, several interesting questions emerge. First, extending the framework to handle continuous sensitive variables presents a natural challenge. While categorical variables allow for a simple matrix inversion during reversion, continuous variables might require solving integral equations. Second, the method can be further generalized to address cases where multiple variables have undergone privacy-preserving transformations. Combining multiple covariates into a single variable offers a straightforward solution for certain scenarios (e.g., combining binary variables $X_1$ and $X2$ into a new variable with four levels). However, the problem becomes significantly more complex when dealing with a mix of categorical and continuous variables.

\newpage

\bibliographystyle{apalike}
\bibliography{reference}

\end{document}